\documentclass{elsart}
\usepackage{graphicx}
\newcommand{\pd}[2]{\frac{\partial #1}{\partial #2}}
\renewcommand{\vec}[1]{{\bf #1}} 
\begin{document} 
\begin{frontmatter}
\title{Granular and Nano-Elasticity}
\author[Eng]{I. Goldhirsch},
\ead{isaac@eng.tau.ac.il}
\author[Physics]{C. Goldenberg}
\ead{chayg@post.tau.ac.il}
\address[Eng]{Department of Fluid Mechanics and Heat Transfer,
Faculty of Engineering\\
Tel-Aviv University, Ramat-Aviv, Tel-Aviv 69978, Israel}
\address[Physics]{School of Physics and Astronomy\\ 
Tel-Aviv University, Ramat-Aviv, Tel-Aviv 69978, Israel}
\date{January 6, 2002}
\begin{abstract} 
  The modeling of the elastic properties of granular or nanoscale systems
  requires the foundations of the theory of elasticity to be revisited, as one
  explores scales at which this theory may no longer hold. The only cases for
  which a microscopic justification of elasticity exists are (nearly) uniformly
  strained lattices. A microscopic theory of elasticity, as well as
  simulations, reveal that standard continuum elasticity applies only at
  sufficiently large scales (typically 100 particle diameters).  Interestingly,
  force chains, which have been observed in experiments on granular systems,
  and attributed to non-elastic effects, are shown to exist in systems composed
  of harmonically interacting constituents. The corresponding stress field,
  which is a continuum mechanical (averaged) entity, exhibits no chain
  structures even at near-microscopic resolutions, but it does reflect
  macroscopic anisotropy, when present.
\end{abstract} 
\end{frontmatter}
%=============================================================================== 
\section{Introduction} 
\label{sec:intro}
%===============================================================================
It is quite surprising that the existing microscopic justification of the
time-honored theory of elasticity, which has been thoroughly researched in a
variety of disciplines, is limited to lattice atomic
configurations~\cite{KittelFeynman}. Classical continuum elasticity theory has
been applied to a large variety of systems, including granular
materials~\cite{GranularElasticity}. In recent years the same theory has been
applied for the description of elastic properties of micro- and nano-scale
systems (e.g.,~\cite{MicroNano}). It is a-priori unclear whether this theory
applies at such small scales.

The study presented below shows that the justification of elastic theory based
on a `microscopic' picture is not entirely straightforward. In fact, we show
that continuum elasticity is limited to sufficiently large scales. One of the
other results presented in this paper is that `force chains' in granular
systems, which have been attributed by some authors to non-elastic
effects~\cite{NonElastic}, are reproduced in simulations of elastic systems, in
good agreement with experiments.

%===============================================================================
\section{Coarse Graining and Constitutive Relations} 
\label{sec:cg}
%===============================================================================
%-------------------------------------------------------------------------------
\subsection{Preliminaries}
\label{sec:prelim}
%-------------------------------------------------------------------------------
Below the term `particles' is taken to mean grains in a granular system or
atoms in a solid. Classical mechanics is assumed throughout this paper.
Following~\cite{Glasser01} define, for a system of particles, the coarse
grained mass and momentum densities, at the position $\vec{r}$ and time $t$, by
\mbox{$\rho({\bf r},t)\equiv \sum_i m_i\phi[{\bf r}-{\bf r}_i(t)]$} and
\mbox{$\vec{p}({\bf r},t)\equiv \sum_i m_i \vec{v}_i(t) \phi[{\bf r}-{\bf
    r}_i(t)]$}, respectively, where $\{{\bf r}_i(t); {\bf v}_i(t);m_i\}$ are
the positions, velocities and masses of the particles, indexed by $i$, and
$\phi(\vec{R})$ is a normalized non-negative coarse graining function (with a
single maximum at $\vec{R}=0$) of width $\lambda$, the coarse graining scale.
The velocity field is defined by \mbox{$\vec{V}({\bf r},t)\equiv \vec{p}({\bf
    r},t)/\rho({\bf r},t)$}. Upon taking the time derivatives of these coarse
grained fields and performing straightforward algebraic
manipulations~\cite{Glasser01} one obtains the equation of continuity, as well
as the momentum conservation equation: \mbox{$\dot{p}_\alpha({\bf r},t)= -
  \frac{\partial}{\partial r_\beta} \left[ \rho({\bf r},t) V_{\alpha}({\bf
      r},t) V_{\beta}({\bf r},t) - \sigma_{\alpha\beta}({\bf r},t)\right]$},
where Greek indices denote Cartesian coordinates. The stress tensor,
$\sigma_{\alpha\beta}$, is composed of a kinetic contribution, irrelevant for
our purposes (we consider quasi-static deformations), and a ``contact''
contribution. Neglecting the former, one obtains:
\begin{equation}
\label{qs_stress}
\sigma_{\alpha\beta}({\bf r},t) =  -\frac{1}{2} \sum_{i,j;i\ne j} 
f_{ij\alpha}(t) { r}_{ij\beta}(t) \int_0^1 ds \phi[{\bf r}-{\bf r}_i(t) + 
s {\bf r}_{ij}(t)]. 
\end{equation} 
where $f_{ij\alpha}(t)$ is the $\alpha$-th component of the force exerted on
particle $i$ by particle $j$ ($j \neq i$) at time $t$ (assuming pairwise
interactions) and $r_{ij\alpha}\equiv r_{i\alpha}-r_{j\alpha}$.
%-------------------------------------------------------------------------------
\subsection{Displacement and Strain}
\label{sec:disp+strain}
%-------------------------------------------------------------------------------
Following elementary continuum mechanics, the velocity of a material particle
whose initial (Lagrangian) coordinate is $\vec{R}$ satisfies:
\mbox{$\vec{V}^{\rm La}(\vec{R},t)=\partial{\vec{u}^{\rm
      La}\left(\vec{R},t\right)}/\partial{t}$}, where $\vec{u}^{\rm
  La}(\vec{R},t)\equiv \vec{r}(\vec{R},t)-\vec{R}$ is the (Lagrangian)
displacement field. Therefore: \mbox{$\vec{u}^{\rm La}(\vec{R},t)=\int_0^t
  \vec{V}^{\rm La}(\vec{R},t')dt'$}.  Using the definitions presented in
Sec.~\ref{sec:prelim}, it follows that:
\begin{equation}
  \label{eq:displacement_la_exact}
\vec{u}^{\rm La}(\vec{R},t) \equiv
\int_0^t \frac{\sum_{i} m_i \vec{v}_i(t')
  \phi[\vec{r}(\vec{R},t')-\vec{r}_i(t')]} 
    {\sum_{j} m_j \phi[\vec{r}(\vec{R},t')-\vec{r}_j(t')]}dt'.
\end{equation}
Noting that $\dot{\vec{u}}_i=\vec{v}_i$, where $\vec{u}_i$ is the displacement
of particle $i$, and invoking integration by parts in
Eq.~(\ref{eq:displacement_la_exact}), one obtains:
\begin{equation}
  \label{eq:displacement_eu_approx}
  \vec{u}(\vec{r},t) \simeq
  \frac{\sum_{i} m_i \vec{u}_i(t)
    \phi[\vec{r}-\vec{r}_i(t)]} 
    {\sum_{j} m_j \phi[\vec{r}-\vec{r}_j(t)]} + 
\mbox{``non-linear terms''},
\end{equation}
where the result has been translated to the Eulerian representation.  The
correction terms can be shown to beget non-linear terms in the constitutive
relations. The linear strain is given by
\mbox{$\epsilon_{\alpha\beta}(\vec{r},t)\equiv \frac{1}{2}\left[\frac{\partial
      u_\alpha(\vec{r},t)}{\partial r_\beta} + \frac{\partial
      u_\beta(\vec{r},t)}{\partial r_\alpha}\right]$}. Note that the exact
displacement and strain obtained here are different from the commonly defined
mean field strain, cf. e.g.~\cite{Bathurst88}. Mean field theories are based on
the assumption that the relative particle displacements are described by the
macroscopic strain, \mbox{$u_{ij\alpha}(\vec{r},t)=\epsilon_{\alpha\beta}
  R_{ij\beta}$}, where $\vec{u}_{ij}\equiv \vec{u}_{i}-\vec{u}_{j}$. A
suggested improvement is the ``best fit'' strain~\cite{Liao97}. These mean
field approaches are inaccurate and often lead to incorrect constitutive
relations~\cite{GoldenbergUP}.

%-------------------------------------------------------------------------------
\subsection{Stress-Strain Relation}
\label{sec:stress_strain}
%-------------------------------------------------------------------------------
The microscopic expressions for the stress [Eq.~(\ref{qs_stress})] and the
strain are not manifestly proportional. Therefore elasticity is not a-priori
obvious.  To see how it still comes about, consider the case of harmonic (and
local) pairwise particle potentials
$U_{ij}=\frac{1}{2}K_{ij}\left(|r_{ij}|-l_{ij}\right)^2$.  The forces are then
given, to linear order in \mbox{$\delta\vec{r}_{ij}\equiv \vec{u}_{ij}$}, by
\mbox{$\vec{f}_{ij}=-K_{ij}\left(\hat{\vec{r}}^0_{ij}\cdot \delta\vec{r}_{ij}
  \right) \hat{\vec{r}}^0_{ij}$}. The stress [Eq.~(\ref{qs_stress})] is then
given by \mbox{$\sigma_{\alpha\beta}^\mathrm{lin}({\bf r},t) = \frac{1}{2}
  \sum_{ij} K_{ij}\hat{r}^0_{ij\gamma} \delta r_{ij\gamma}\hat{r}^0_{ij\alpha}
  {r}^0_{ij\beta} \phi({\bf r}-{\bf r}^0_i)$}, up to nonlinear terms in
$\left\{\delta\vec{r}_{ij}\right\}$. The superscript $0$ denotes the reference
configuration.

Consider a volume $\Omega$ which is much larger than the coarse graining scale,
$\lambda$, and let $\vec{r}$ be an interior point of $\Omega$ which is `far'
from its boundary.  Let upper case Latin indices denote the particles in the
exterior of $\Omega$ which interact with particles inside $\Omega$.  Since the
considered system is linear, there exists a Green's function $\bf G$ such that
\mbox{$u_{i\alpha}=G_{i\alpha J\beta}u_{J\beta}$} for $i\in \Omega$. Hence,
\mbox{$u_{ij\alpha}=\left(G_{i\alpha J\beta}-G_{j\alpha
      J\beta}\right)u_{J\beta}$}. When all $\vec{u}_J$ are equal (rigid
translation), \mbox{$\vec{u}_{ij}=0$}. Hence,
\mbox{$u_{ij\alpha}=\left(G_{i\alpha J\beta}-G_{j\alpha
      J\beta}\right)\left[u_{J\beta}-u_\beta(\vec{r})\right]$}. Define
\mbox{$L_{ij\alpha J\beta}\equiv G_{i\alpha J\beta}-G_{j\alpha J\beta}$}.  One
then obtains \mbox{$u_{ij\alpha}=L_{ij\alpha J\beta}
  \left[u_\beta(\vec{r}_J)-u_\beta(\vec{r})\right] +L_{ij\alpha J\beta}
  \left[u_{J\beta}-u_\beta(\vec{r}_J)\right]$}, where
\mbox{$u_{J\beta}-u_\beta(\vec{r}_J)$} is a fluctuating displacement. The sum
over $J$ in the second term can be shown to be subdominant when the linear size
of the volume $\Omega$ greatly exceeds the coarse graining scale. Furthermore,
to lowest order in a gradient expansion,
\mbox{$u_\beta(\vec{r}_J)-u_\beta(\vec{r})\simeq
  \pd{u_{\beta}(\vec{r})}{r_{\gamma}}\left(r_{J\gamma}-r_\gamma\right) $}.
Substituting this result in $\mbox{\boldmath{$\sigma$}}^\mathrm{lin}$, and
employing rotational symmetry, one obtains:
\begin{equation}
\label{stress_strain_linear}
\sigma_{\alpha\beta}(\vec{r}) \simeq  \frac{1}{2}
\epsilon_{\mu\nu}(\vec{r}) 
\sum_{ij} K_{ij} L_{ij\gamma J\mu}
\left(r^0_{J\nu}-r_\nu\right)
\hat{r}^0_{ij\alpha}{r}^0_{ij\beta}  \hat{r}^0_{ij\gamma} 
\phi({\bf r}-{\bf r}^0_i).
\end{equation} 
Thus linear elasticity is valid when $\left| \lambda \nabla_{\alpha}
  \nabla_{\beta} \vec{u} \right|\ll 1$ and
$\left\|\mbox{\boldmath{$\epsilon$}}\right\|\ll 1$. Note that the elastic
moduli depend, in principle, on the position as well as the resolution (through
the coarse graining function $\phi$).

%-------------------------------------------------------------------------------
\section{Numerical Results}
\label{sec:numerical}
\begin{figure}[!ht]
\begin{center}
\begin{tabular}{cc}
  \includegraphics[width=2.75in]{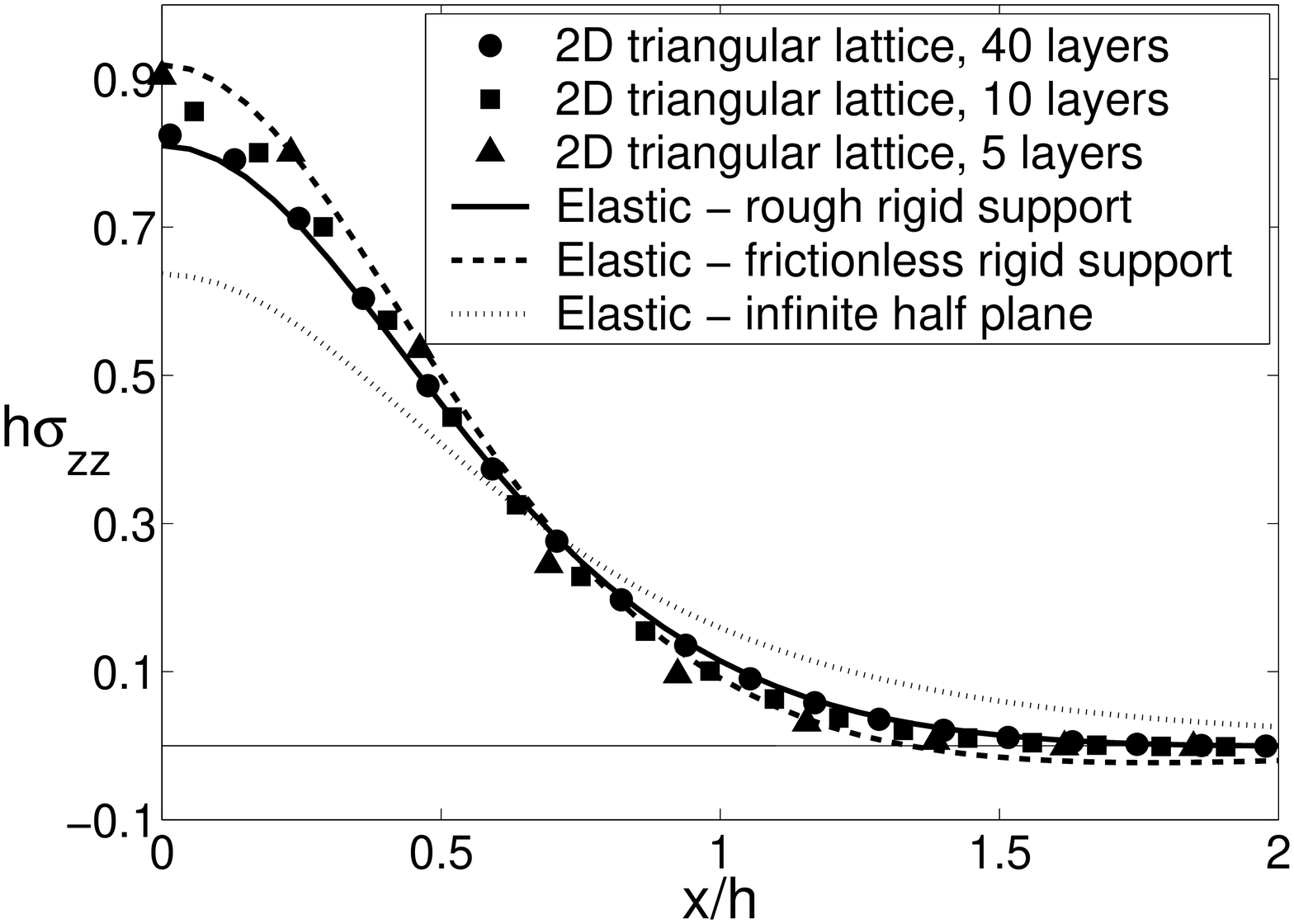}&
  \includegraphics[width=2.75in]{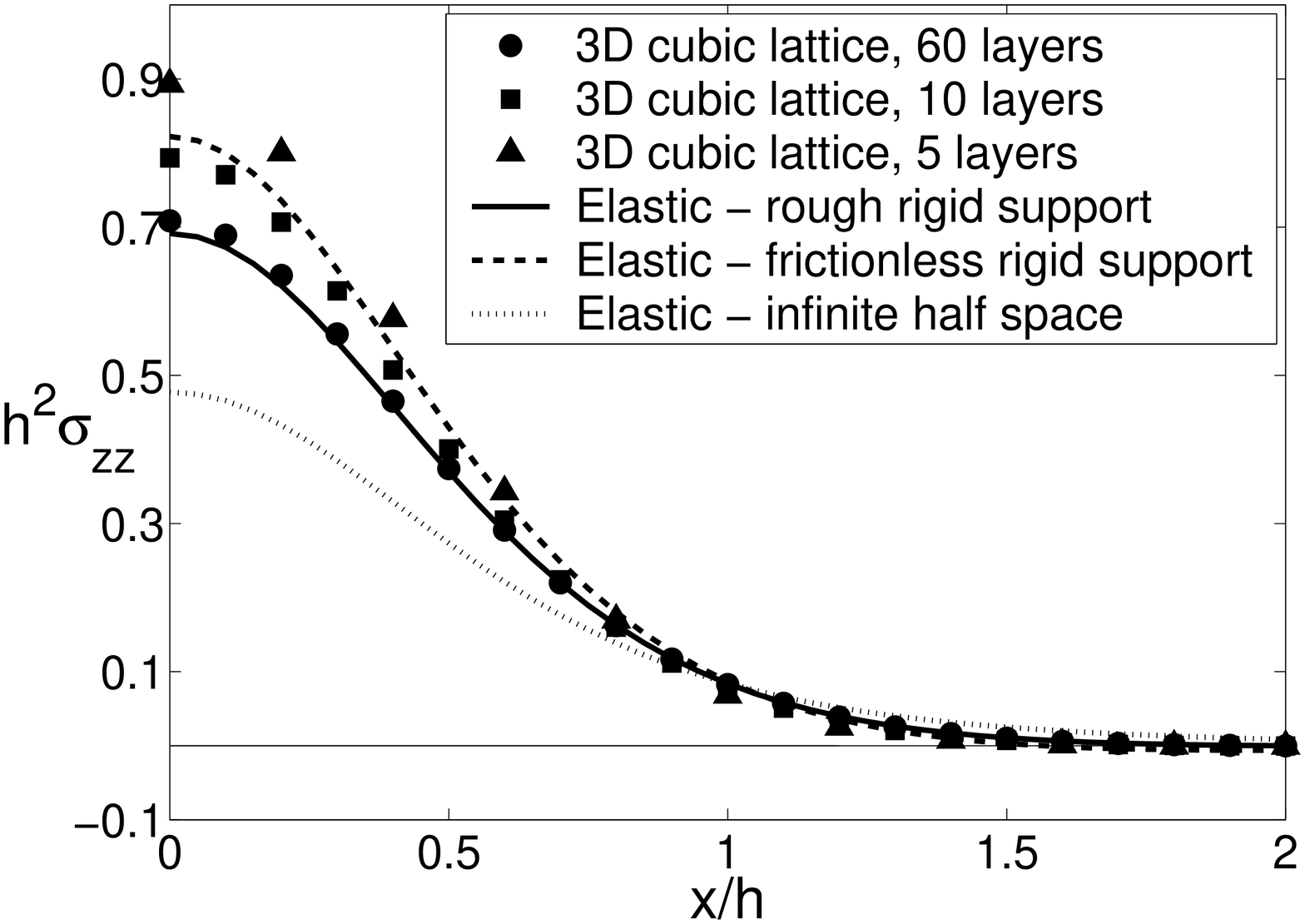}
\end{tabular}
\end{center}
\caption{The vertical stress at the bottom of a 2D triangular lattice (left)
  and a 3D cubic lattice (right), compared to continuum elastic solutions. The
  results are scaled by the slab height $h$.}
\label{fig:Crossover}
\end{figure}

%-------------------------------------------------------------------------------
In order to study the crossover from microelasticity to
macroelasticity~\cite{Goldenberg01}, we have considered a two dimensional (2D)
rectangular slab, composed of uniform disks (of diameter $d$) whose centers are
positioned on a triangular lattice, with nearest neighbors coupled by uniform
linear springs (of rest length $d$), and a three dimensional (3D) slab,
composed of uniform spheres (of diameter $d$) with springs coupling
nearest-neighbors (spring constant $K_1$) and next-nearest-neighbors ($K_2$).
Both systems can be shown to correspond to {\em isotropic} elastic media in the
continuum limit~\cite{KittelFeynman} (for the 3D system, only if $K_1=K_2$).
The particles at the bottom layer are attached to a rigid floor, and a downward
force is applied to the central particle in the top layer.  This is intended
for comparison with 2D~\cite{ForceChainsExp} and 3D~\cite{Clement} granular
experiments (even though these models do not directly describe the interaction
of granular particles and are primarily intended for studying the crossover
between microelasticity and macroelasticity, they closely reproduce some of the
experimental results).  Fig.~\ref{fig:Crossover} presents a comparison between
the vertical stress exerted on the floor of the system for different slab
heights (numbers of layers of particles) and continuum elastic solutions, for
both systems ($K_1=K_2$ for the 3D system).  The convergence to the appropriate
(rough rigid support) elastic stress distribution for a sufficient number of
layers is evident.

\begin{figure}[!ht]
\begin{center}
\begin{tabular}{cc}
  \includegraphics[width=2.75in]{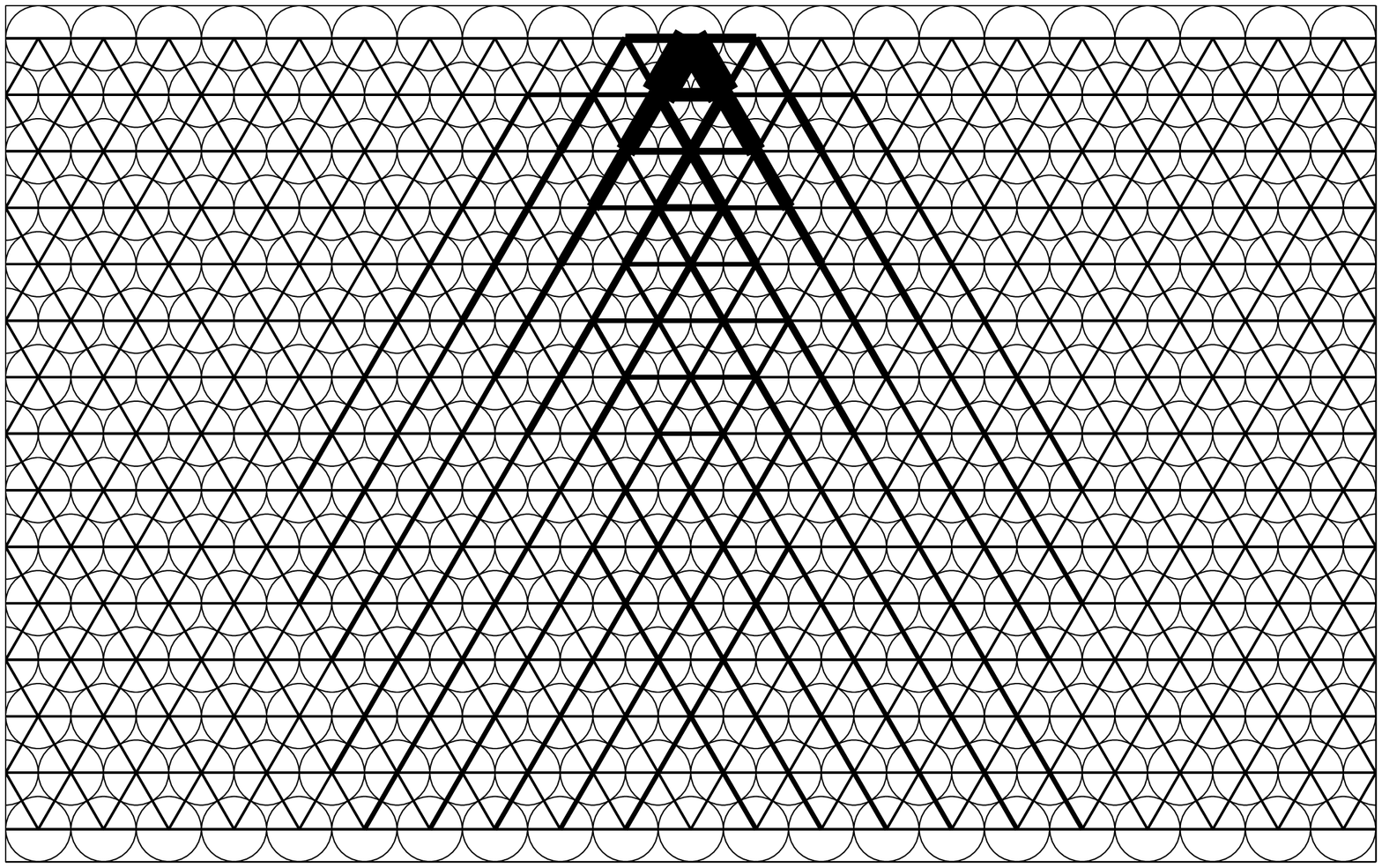}&
  \includegraphics[width=2.75in]{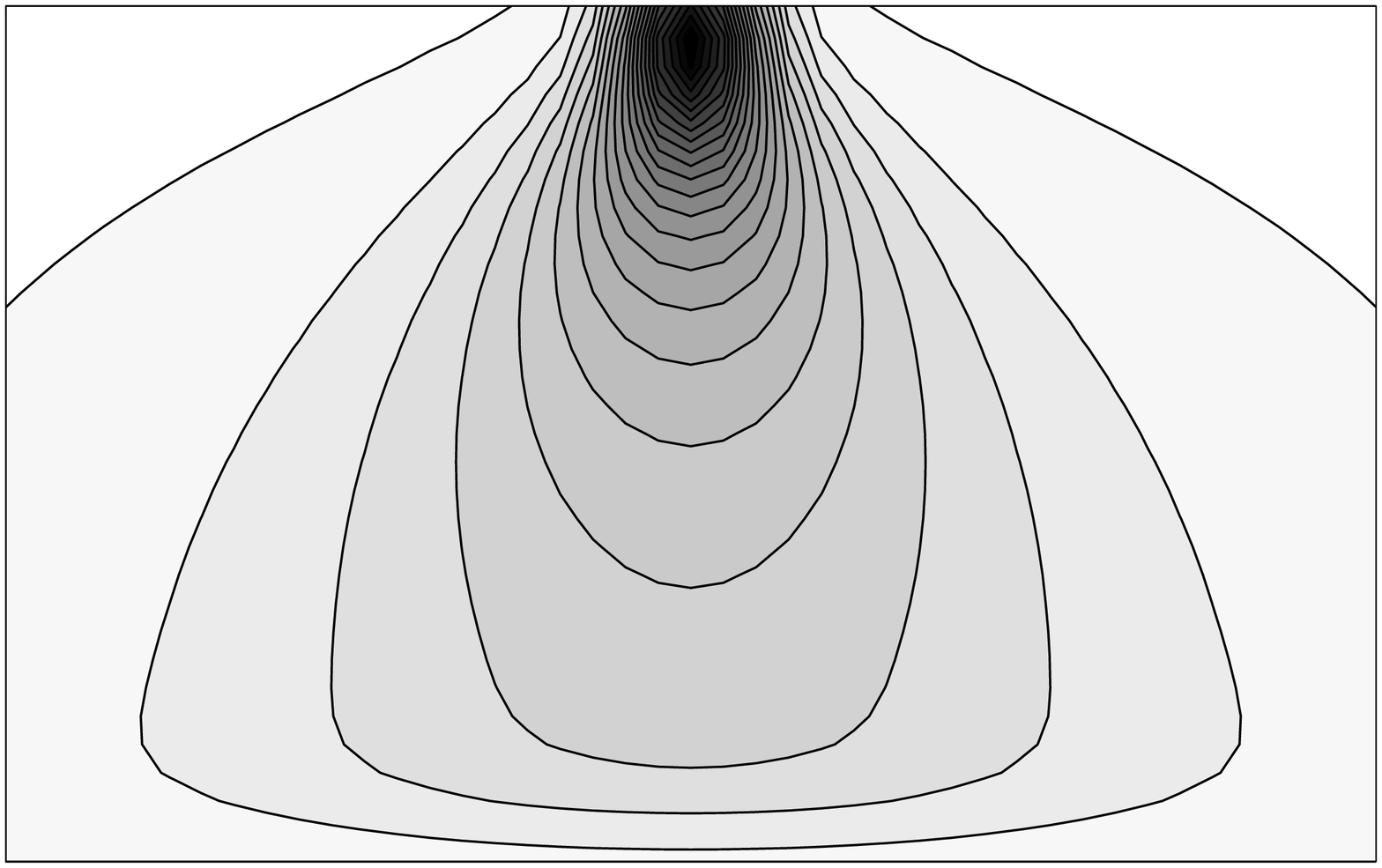}
\end{tabular}
\end{center}
\caption{Left: Force chains in a 2D triangular lattice. A vertical force is applied
  to the center particle in the top layer. The line widths are proportional to
  the forces. Only the central region of the lattice is shown (the lattice
  comprises $15\times41$ particles). Right: Contour plot of the vertical stress
  component, $\sigma_{zz}$, in the same region.}
\label{fig:lat2D_forces+stress}
\end{figure}

The observation of ``force chains'' in experiments~\cite{ForceChainsExp} and
simulations~\cite{ForceChainsTheory} has been interpreted as evidence for
non-elastic effects~\cite{NonElastic}. However, in a plot of the interparticle
forces in the 2D system described above (Fig.~\ref{fig:lat2D_forces+stress},
left) one can identify ``force chains'' though the microscopic interactions are
strictly linear, and the large scale behavior corresponds to isotropic
elasticity.  The corresponding distribution of interparticle
forces~\cite{Goldenberg01} closely resembles the experimental findings of Geng
et al.~\cite{ForceChainsExp}.  The $zz$ component of the stress field,
calculated using Eq.~(\ref{qs_stress}) with a Gaussian coarse graining
function~\cite{Glasser01} of width $d$ (i.e., with fine resolution), is shown
in Fig.~\ref{fig:lat2D_forces+stress}, on the right.  The force chains are not
evident any more. The force chains result from the microscopic anisotropy
dictated by the fact that a single particle is never subject to an isotropic
force distribution.  However, even at small, but finite, spatial resolution,
the calculation of the stress tensor inherently involves averaging over the
forces acting on the particles, so that this anisotropy does not appear in the
stress field.  Similar results~\cite{Goldenberg01} are obtained for random 2D
systems.

A more realistic model with unilateral (``one-sided'') springs, active only
upon compression, yields results which are even closer to the
experiments~\cite{Goldenberg01}. In this case, some of the springs are
disconnected (as obtained in~\cite{Luding97} for a pile geometry) when the
external force is applied. In particular, the horizontal springs in a
triangular region below the point of application of the force are severed, and
the corresponding macroscopic properties are anisotropic. The resulting stress
distribution~\cite{Goldenberg01} is similar to that predicted by hyperbolic
models of force propagation~\cite{NonElastic}, but it is a result of an elastic
model.

\begin{ack}
\label{sec:ack}
Support from the Israel Science Foundation, grants no. 39/98 and 53/01, is
gratefully acknowledged.
\end{ack}

\end{document}